\documentclass[10pt]{article}
\usepackage{geometry}
\usepackage[utf8]{inputenc}
\usepackage{amsmath}
\usepackage{amsfonts}
\usepackage{amssymb}
\usepackage{makeidx}
\usepackage{graphicx}
\usepackage{verbatim}
\usepackage{wrapfig}
\usepackage{xcolor}
\usepackage{multicol}
\usepackage{comment}
\usepackage{setspace}
\usepackage{soul}
\usepackage{makeidx}
\usepackage{braket}
\usepackage[font=small,labelfont=it]{caption}
\usepackage[font=scriptsize]{caption}
\usepackage{blindtext}

\def\Z{\mathbb Z}

\def\R{\mathbb R}

\newcommand{\De}{{\Delta\vphantom{\big|}}}
\newcommand{\Xe}{{X\vphantom{\big|}}}

\newcommand{\E}{{\scriptscriptstyle E}}

\newcommand{\cD}{{\cal D}}

\renewcommand{\>}{{\rangle}}

\newcommand{\sfrac}[2]{{\textstyle\frac{#1}{#2}}}
\newcommand{\half}{{\sfrac12}}
\newcommand{\pa}{\partial}

\newcommand{\res}{{\mathrm{res}}}

\newcommand{\im}{{\mathrm{i}}}
\newcommand{\mi}{{\mathrm{i}}}

\newcommand{\diff}{{\mathrm{d}}}

\newcommand{\gs}{g_s}
\newcommand{\gl}{g_{\ell}}

\newcommand{\beq}{\begin{equation}}
\newcommand{\eeq}{\end{equation}}
\newcommand{\eq}{\end{equation}}
\newcommand{\bea}{\begin{eqnarray}}
\newcommand{\eea}{\end{eqnarray}}

\renewcommand{\and}{{\quad{\rm and}\quad}}
\newcommand{\und}{{\qquad{\rm and}\qquad}}
\renewcommand{\=}{\ =\ }

\begin{document}
\begin{center}
{\LARGE \bf 
Integrability, intertwiners and non-linear algebras \\[4pt] 
in Calogero models
\\
}
\vspace{6mm}
{\Large Francisca Carrillo--Morales$^a$, Francisco Correa$^{a}$ and Olaf Lechtenfeld$^b$}
\\[6mm]
\noindent ${}^a${\em 
Instituto de Ciencias F\'isicas y Matem\'aticas\\
Universidad Austral de Chile, Casilla 567, Valdivia, Chile}\\[3mm]
\noindent ${}^b${\em
Institut f\"ur Theoretische Physik and Riemann Center for Geometry and Physics\\
 Leibniz Universit\"at Hannover, Appelstrasse 2, 30167 Hannover, Germany}
\vspace{12mm}
\end{center}

\begin{abstract} \noindent
For the rational quantum Calogero systems of type $A_1{\oplus}A_2$, $AD_3$ and $BC_3$, 
we explicitly present complete sets of independent conserved charges and their nonlinear algebras.
Using intertwining (or shift) operators, we include the extra `odd' charges appearing for integral couplings.
Formul\ae\ for the energy eigenstates are used to tabulate the low-level wave functions.
\end{abstract}

\section{Introduction and summary}

There is a vast literature on quantum Calogero (or Calogero--Moser--Sutherland) models as
a paradigm for a superintegrable system with a finite number of degrees of freedom
\cite{calogero,woj}. 
Since these systems admit an analytic computation of basically every detail, 
a variety of rich mathematical structures has been uncovered
(for reviews, see e.g.~\cite{OlPe,poly2}; for a recent generalization,
see \cite{poly3}).
Nevertheless, some of the latter have not been displayed very explicitly
or are hidden in mathematical literature difficult to penetrate for most physicists.

Among these aspects is the nonlinear algebra formed by the $2N{-}1$ (or, for integral coupling, $2N$)
independent conserved charges in the case of the rational rank-$N$ Calogero model~\cite{kuz,CLP}.
A related feature is the form of the intertwiners (or shift operators). These relate the simultaneous
eigenstates of the $N$~Liouville charges at integer-spaced coupling values, 
thereby providing an alternative access to those states for integral couplings 
and allowing for the $2N$th extra charge~$Q$.

For this reason, 
we provide completely explicit formul\ae\ for the rank-3 rational quantum Calogero models,
based on the Coxeter reflection groups $A_1{\oplus}A_2$, $AD_3$ and $BC_3$ (but not $H_3$),
for the nonlinear algebras, the intertwiners and the energy eigenstates.
In contrast to the customary $A_1{\oplus}A_2$ model (describing three nonrelativistic unit-mass 
particles on the infinite line and interacting pairwise via an inverse-square two-body potential),
the $AD_3$ and $BC_3$ models have been investigated much less.
Yet, even the $A_1{\oplus}A_2$ case has not been fully analyzed:
The center-of-mass sector (associated to the $A_1$ part) is usually free, 
as one imposes translational invariance as a physical prejudice.
However, nothing prevents one from giving it its own (external) inverse-square potential.
In fact, such is completely natural in the bigger scope, and here we present all details
for this generalized three-particle system as well.

The organization of the paper is as follows. In the remainder of this introduction, we introduce 
the Dunkl operators~\cite{dunkl,roesler} as our major tool for the construction of conserved charges,
then review how the conformal algebra can be used to extend the set of Liouville charges to 
an infinite (but functionally dependent) set of higher integrals of motion, and finally present
the concept of intertwining operators and how they give rise to additional conserved charges
in the special situation of integral coupling(s). 
Sections 2, 3 and 4 then treat the cases of $A_1{\oplus}A_2$, $AD_3$ and $BC_3$ in turn, 
showing in each instance a full set of independent conserved charges, 
a formula for the energy eigenstates, the concrete form of the basic intertwiners,
and finally the nonlinear algebra of all charges including the extra ones at integral coupling.
Some lengthy expressions and lists of low-lying energy eigenstates are delegated to an appendix.

\subsection{Liouville charges}
Let us consider a set $R^+$ of positive roots $\alpha$ for a Coxeter group $W$ of reflections $s_\alpha$ 
in $\mathbb{R}^N\ni x$. Then, the Dunkl operators associated to the standard Cartesian basis $\{e_i\}$
with $i=1,\ldots,N$ are given by
\begin{equation}\label{Dk}
\mathcal{D}_i\=\partial_i-\sum_{\alpha\in R^+}\dfrac{g_{\alpha}\alpha_i}{(\alpha,x)}s_\alpha
\qquad\text{with}\quad (\alpha,x)=\alpha(x) \and \alpha_i=(\alpha,e_i)=\alpha(e_i)\ ,
\end{equation}
where $\pa_i=\pa/\pa x^i$ for $x=e_i x^i$, and we canonically identify $\R^N$ with its dual.
The real coupling constants $g_\alpha$ depend only on the Weyl orbit of~$\alpha$,
so we shall encounter at most two values, either $\gs$ for the short roots and $\gl$ for the long roots, 
or a single coupling $g$ in the simply-laced case.
A key role is played by the Weyl-invariant polynomials $\sigma_k(x)$ of degree~$k$,
because the restriction ``$\res$'' of $\sigma_k(\mathcal{D})$ for $\mathcal{D}\equiv\{\mathcal{D}_i\}$ 
to Weyl-invariant functions yields constants of motion~$I_k$ known as Liouville integrals,
for any (generalized) Calogero model. Since the Dunkl operators mutually commute~\cite{dunkl,roesler}, 
the Liouville integrals also commute with one another. 
The $N$ lowest-order such polynomials will provide $N$ functionally independent Liouville charges.

A universal Weyl-invariant polynomial is $\sigma_2(x)=(x,x)=:r^2$. 
The corresponding integral $I_2$ is (minus twice) the Hamiltonian of the system, 
\begin{equation}\label{H}
H\=-\tfrac12\res(\mathcal{D},\mathcal{D})
\=\tfrac{1}{2}\sum_{i=1}^N p_i^2+\sum_{\alpha\in R^+}\frac{g_\alpha(g_\alpha{-}1)(\alpha,\alpha)}{2\,(\alpha,x)^2}
\qquad\text{with}\quad p_i = -\im\pa_i\ ,
\end{equation}
and is invariant under $g_\alpha \to 1{-}g_\alpha$ for any $\alpha$.
For $g_\alpha=0$ or $g_\alpha=1$, the (free) Liouville charges are simply given by
\begin{equation}
I_k \= \sigma_k(p)\ .
\end{equation}

\subsection{Energy spectrum}
The energy spectrum is continuous with $E\ge0$ but highly degenerate.
The $H$ eigenstates may be labelled by the energy~$E$ and 
by $N{-}1$ additional quantum numbers $\ell_k$ 
corresponding to the Weyl-invariant polynomials $\sigma_k$ (other than $\sigma_2$)
and combining in the generalized angular momentum 
\begin{equation}
q \= \sum_\alpha g_\alpha + {\sum_{\{k\}}}' k\,\ell_k
\end{equation}
where the prime indicates leaving out $k{=}2$.
They are conveniently found by separating the Schr\"odinger equation in spherical coordinates
$(r,\vec{\theta})$, yielding a basis of eigenfunctions
\begin{equation}
\Psi^{(\{g_\alpha\})}_{\E,\{\ell_k\}}(x) \= 
r^{-(N-2)/2}\,J_{q+(N-2)/2}({\scriptstyle\sqrt{2E}}\,r)\ v^{(\{g_\alpha\})}_{\{\ell_k\}}(\vec{\theta})
\end{equation}
with a Bessel-type radial dependence. The angular part of the wave function is best expressed as
\begin{equation}
v^{(\{g_\alpha\})}_{\{\ell_k\}}(\vec{\theta}) \= r^{-q}\De^g\,h^{(\{g_\alpha\})}_{\{\ell_k\}}(x) 
\qquad\text{with}\quad
\De^g \= \prod_{\alpha} (\alpha,x)^{g_\alpha}
\end{equation}
and a Dunkl-deformed harmonic polynomial~\footnote{
It is annihilated by $\tilde{H}=\De^{-g}H\,\De^g=-\tfrac12\sum_i\partial_i^2+O(g)$.}
\begin{equation}
h^{(\{g_\alpha\})}_{\{\ell_k\}}(x) \= 
r^{N-2+2q}\,\Bigl\{{\prod_{\{k\}}}' \sigma_k(\mathcal{D})^{\ell_k}\Bigr\}\,r^{-N+2-2\sum_\alpha g_\alpha}
\end{equation}
of degree $\sum^\prime_{\{k\}} k\,\ell_k$.
It may be noted that these states generally are not eigenstates of the other Liouville charges
but can be linearly combined to jointly diagonalize all of them.

\subsection{Conformal algebra}
The Hamiltonian~\eqref{H} and the two operators
\begin{equation}\label{confOp}
D=\tfrac{1}{2}\sum_{i=1}^N (x^ip_i+p_ix^i) \und
K=\tfrac{1}{2}\sum_{i=1}^N (x^i)^2 = \half(x,x)
\end{equation}
form the basis of a conformal algebra $sl(2,\mathbb{R})$, 
\begin{equation}
[D,H]=2\mi H\ ,\qquad[D,K]=-2\mi K\ ,\qquad[K,H]=\mi D\ ,
\end{equation}
where $D$ generates the scale transformations (or dilatations) and $K$ the special conformal transformations.

This dynamical symmetry is enhanced by extending $H$ to the entire set $\{I_k\}$ of Liouville charges, 
which produces an infinite quadratic algebra~\cite{kuz,CLP}.
The dilatation operator yields a $\Z$ grading, but each commutation with $K$ provides a new layer of operators,
starting with
\begin{equation}\label{J_k}
\tfrac{1}{\mi}[K,I_k]= k J_k
\end{equation}
satisfying the commutation relations
\begin{equation}\label{Jk2}
\tfrac{1}{\mi}[D,J_k]=(k{-}2)J_k \und \tfrac{1}{\mi}[H,J_k]=-I_k\ .
\end{equation}
Note that $J_1=\sum_i x^i$ and $J_2=D$.
There are two obvious ways to build further integrals of motion, which will not be in involution however.
The first one admits explicit time dependence,
\begin{equation}
\bar{J}_k \ :=\ J_k - t\,I_k \qquad\Rightarrow\qquad
\sfrac{\diff}{\diff t} \bar{J}_k \= \partial_t \bar{J}_k +\im[H,\bar{J}_k] \= 0\ .
\end{equation}
The second one forms the antisymmetric combinations
\begin{equation}\label{L_kl}
L_{k,\ell}\ :=\ \tfrac{1}{2}(I_k J_{\ell}+J_{\ell} I_k)-\tfrac{1}{2}(I_{\ell} J_k+J_k I_{\ell})
\qquad\Rightarrow\qquad [H,L_{k,\ell}] \= -\tfrac{\im}{2}(I_k I_\ell+I_\ell I_k -I_\ell I_k - I_k I_\ell) \= 0\ .
\end{equation}

The $\bar{J}_k$ or the $L_{k,\ell}$ form overcomplete sets of constants of motion, and there exist
many options for a functionally independent complete subset. 
Here, we choose
\begin{align}\label{F_l}
F_k\ :=\ L_{2,k} \= \{H,J_k\} -\half\{I_k,D\}
\end{align}
with the same $N$ lowest values for $k$ determined by the first Weyl-invariant polynomials~$\sigma_k$.
Since $F_2\equiv0$ by definition, these provide $N{-}1$ additional integrals of motion, 
revealing the complete superintegrability of the rational quantum Calogero model.
Our choice of extra integrals bears a close relation to the Casimir element of the conformal algebra,
\begin{equation}\label{Cas}
C\=KH+HK-\tfrac{1}{2}D^2\ ,
\end{equation}
which generates the $F_k$ directly from the $I_k$,
\begin{equation}
\tfrac{1}{\mi}[C,I_k] \= k\,F_k\ .
\end{equation}
The quadratic algebra spanned by the $I_k$ and $F_k$ has been presented in \cite{CLP}.

\subsection{Intertwining operators}
The previous results were obtained for generic real couplings $g_\alpha$. 
For integer coupling values, there appears one additional independent constant of motion, 
due to the invariance of~$H$ under $g\to1{-}g$ and the existence of intertwining (or shift) operators $M(g)$.
The latter are constructed from Weyl {\it anti\/}-invariant polynomials~$\tau_m(x)$, 
again by replacing the arguments $x^i$ with the Dunkl operators~$\mathcal{D}_i$ 
and restricting the result to Weyl-symmetric functions, as we did for the construction of the Liouville integrals.
The degree $m$ of those polynomials and the number of independent ones 
depend on the root system under consideration. In this sense, intertwining operators have been studied under
this approach in angular and trigonometric Calogero models \cite{cl1,cl2,cl3}.

Let us be more concrete for the simply-laced situation, $g_\alpha=g$.
Any intertwiner $M(g)$ establishes a relation between the Liouville integrals at integrally shifted couplings, 
\begin{equation}\label{Intrel}
M(g)\,I_k(g)\=I_k(g{+}1)\,M(g) \und
M(1{-}g)\,I_k(g)\=I_k(g{-}1)\,M(1{-}g)\ ,
\end{equation}
and hence transports eigenstates of $I_k(g)$ to eigenstates of $I_k(g{+}1)$ (or zero).
In particular,
\begin{equation}
M(g)\,\Psi^{(g)}_{\E,\{\ell_k\}}(x) \= \sum_{\{k\}} c_{\{\ell_k\}}^{\{\ell'_k\}}(g)\,\Psi^{(g+1)}_{\E,\{\ell'_k\}}(x)\ ,
\end{equation}
with some coefficients $c_{\{\ell_k\}}^{\{\ell'_k\}}(g)$.
In this way, simultaneous $I_k$ eigenstates at integer coupling can be obtained from free eigenstates 
by a successive application of shift operators.
As another consequence, by shifting the coupling up and then down again, the operator $M(-g)M(g) $
commutes with all Liouville charges,
\begin{equation}
[M(-g)M(g),I_k(g)]\=0\ ,
\end{equation}
but it is not a new integral of motion since it is expressed in terms of them,
\begin{equation}\label{R(g)}
M(-g)M(g)\= \mathcal{R}(I(g)) \qquad\textrm{for}\quad I=\{I_k\}\ .
\end{equation}
This polynomial of the $I_k$ must not depend on $g$ explicitly (take all $(\alpha,x)\to\infty$).
Therefore, one may easily compute it from the free case $g{=}0$ ,
\begin{equation}
\mathcal{R}(I) \= M(0)^2\ .
\end{equation}

A novel feature appears for integral values of the coupling, say $g=2,3,4,\ldots$.
Shifting it all the way from $1{-}g$ to $g$, the combined intertwiner
\begin{equation}\label{Q(g)}
Q(g)\=M(g{-}1)M(g{-}2)\cdots M(1)M(0)M(-1)\cdots M(2{-}g)M(1{-}g)
\end{equation}
also commutes with all Liouville charges but, as a product of an odd number of intertwiners,
it is functionally independent. Only its square belongs to the ring of Liouville charges,
\begin{equation}
Q(g)^2 \= \mathcal{R}(I(g))^{2g-1}\ .
\end{equation}
Adjoining $Q(g)$ to the $2N{-}1$ conserved quantities $\{I_k,F_k\}$ provides a $\Z_2$ grading and 
makes our model analytically integrable, with $2N$ independent integrals of motion.
We suspect other $Q$ intertwiners based on different shift operators~$M$ to be functionally dependent.
The full set of commutators (also with the $F_\ell$) is given in~\cite{CLP} 
for the $A_{n-1}\oplus A_1$ root system.

In the non-simply-laced case, one expects to find polynomials $\tau'_{m'}$ antisymmetric under
short-root reflections but symmetric under long-root ones, as well as polynomials $\tau_m$ 
with the opposite behavior. Inserting Dunkl operators as arguments and performing the symmetric restriction,
we produce $I_k$ intertwiners $M_s(g_s,g_\ell)$ and $M_\ell(g_s,g_\ell)$, which shift by unity only one coupling
but not the other. 
Also here, those shift operators allow one to build the joint $I_k$ eigenstates for $(g_s,g_\ell)\in\Z\times\Z$
by repeated application on the free eigenstates.
Since we can independently ``wrap'' from $1{-}g$ to $g$ for the short roots or for the long roots, 
there exist two grading operators, $Q_s$ and $Q_\ell$, 
which we expect to be functionally independent of one another.

\newpage

\section{The $\bf{A_1\oplus A_2}$ model}
\subsection{Integrals of motion}
This is the traditional rational three-particle Calogero model, 
enhanced by an external inverse-square potential for the center of mass.
It is reducible because the center-of-mass coordinate and momentum,
\begin{equation}
X \= \sfrac13(x^1+x^2+x^3)  \und P \= p_1+p_2+p_3 \ ,
\end{equation}
can be separated from the other degrees of freedom.
Therefore, we may introduce two coupling constants, say $g$ and $g'$.
The Hamiltonian of the system (\ref{H}) is given by ($i,j=1,2,3$)
\begin{equation}\label{HA}
\begin{aligned}
H &\=\tfrac{1}{2}\sum_i p_i^2
\ +\ \sum_{i<j} \frac{g(g{-}1)}{(x^i{-}x^j)^2}
\ +\ \frac{3\,g'(g'{-}1)}{2(x^1{+}x^2{+}x^3)^2} \\
&\= \tfrac16 P^2\ +\ \frac{g'(g'{-}1)}{6\,X^2}
\ +\ \tfrac16\sum_{i<j}(p_i{-}p_j)^2\ +\ \sum_{i<j} \frac{g(g{-}1)}{(x^i{-}x^j)^2}
\= H_1\ +\ H_2\ .
\end{aligned}
\end{equation}
One can choose the positive roots as
\begin{equation}
\mathcal{R}_+\=\{e_1{-}e_2,\hspace{0.2cm} e_1{-}e_3, \hspace{0.2cm} e_2{-}e_3, \hspace{0.2cm} e_1{+}e_2{+}e_3 \},
\end{equation} 
such that the Dunkl operators (\ref{Dk}) read 
\begin{equation}
\mathcal{D}_i\=\partial_i\ -\ \sum_{j(\neq i)}\frac{g}{x^i{-}x^j}s_{i-j}
\ -\ \frac{g'}{x^1{+}x^2{+}x^3}s_0\ ,
\end{equation}
where the $s_{i-j}$ are the two-particle permutation operators,
\begin{equation} \label{si-j}
\begin{aligned}
s_{1-2}&: \ (x^1,x^2,x^3)\ \mapsto\ (x^2,x^1,x^3)\ , \\
s_{1-3}&: \ (x^1,x^2,x^3)\ \mapsto\ (x^3,x^2,x^1)\ , \\
s_{2-3}&: \ (x^1,x^2,x^3)\ \mapsto\ (x^1,x^3,x^2)\ , \\
\textrm{and}\quad s_0&: \ (x^1,x^2,x^3)\ \mapsto\ (x^1,x^2,x^3)-2X(1,1,1)\ .
\end{aligned}
\end{equation}
The lowest three Weyl-invariant polynomials are
\begin{equation} \label{polyA}
\begin{aligned}
\sigma_2(x) &\= (x^1)^2+(x^2)^2+(x^3)^2 \ ,\\
\tilde{\sigma}_2(x) &\= (x^1{-}x^2)^2+(x^2{-}x^3)^2+(x^3{-}x^1)^2\ ,\\
\tilde{\sigma}_3(x) &\= (x^1{+}x^2{-}2x^3)(x^2{+}x^3{-}2x^1)(x^3{+}x^1{-}2x^2)\ ,
\end{aligned}
\end{equation}
where the center-of-mass coordinate is only contained in~$\sigma_2$.

In this basis, the first three Liouville integrals read
\begin{equation}\label{intsmotionA}
\begin{aligned}
I_2&\= - {\rm res}  \bigl(\mathcal{D}_1^{2}{+}\mathcal{D}_2^{2}{+}\mathcal{D}_3^{2}\bigr) \=2\,H\ ,\\
\tilde{I}_2&\=  -{\rm res} \bigl(
(\mathcal{D}_1{-}\mathcal{D}_2)^2+(\mathcal{D}_2{-}\mathcal{D}_3)^2+(\mathcal{D}_3{-}\mathcal{D}_1)^2\bigr)
\=6\,H_2\ ,\\
\tilde{I}_3&\=\ \im\,{\rm res}  \bigl(
(\mathcal{D}_1{+}\mathcal{D}_2{-}2\mathcal{D}_3)(\mathcal{D}_2{+}\mathcal{D}_3{-}2\mathcal{D}_1)(\mathcal{D}_3{+}\mathcal{D}_1{-}2\mathcal{D}_2) \bigr) \\
&\=\displaystyle \prod_{i=1}^3(P{-}3p_i)-9 g(g{-}1)
\left(\frac{(P{-}3p_3)}{(x^1{-}x^2)^2}+\frac{(P{-}3p_2)}{(x^3{-}x^1)^2}+\frac{(P{-}3p_1)}{(x^2{-}x^3)^2}\right)\ ,
\end{aligned}
\end{equation}
and they are functionally independent and in involution,
\begin{equation}
[I_2,\tilde{I}_2]=[I_2,\tilde{I}_3]=[\tilde{I}_2,\tilde{I}_3]=0\ .
\end{equation}
The two lowest additional integrals of motion, which are not in involution, are 
\begin{equation}
\tilde{F}_2 \= \{H,\tilde{J}_2\} - \half\{\tilde{I}_2,D\} \und
\tilde{F}_3 \= \{H,\tilde{J}_3\}-\ \half\{\tilde{I}_3,D\}\ ,\\
\end{equation}
where
\begin{equation}
\begin{aligned}
\tilde{J}_2 &\=-\tfrac{3}{2}\sum_{i=1}^3 \{X{-}x^i,p_i\} \ ,\\
\tilde{J}_3 &\=\sum_{i=1}^3(P{-}3p_i)(X{-}x^i)(P{-}3p_i) -9g(g{-}1)
\left(\frac{X{-}x^1}{(x^2{-}x^3)^2}+\frac{X{-}x^2}{(x^3{-}x^1)^2}+\frac{X{-}x^3}{(x^1{-}x^2)^2}\right)\ .
\end{aligned}
\end{equation}

\subsection{Energy eigenstates}
The Hamiltonian eigenfunctions are 
for this case are 
\begin{equation}
\Psi^{(g,g')}_{\E,\ell_2,\ell_3}(x) \ \equiv\ \<x \mid \ell_2,\ell_3\>_{g,g'} \=
j_q({\scriptstyle\sqrt{2E}}\,r)\,r^{-q}\De^g\, \Xe^{g'}\,
h^{(g,g')}_{\ell_2,\ell_3}(x) 
\qquad\textrm{with}\qquad q=3g +g' +2\ell_2+3\ell_3\ ,
\end{equation}
where  $\Delta=(x^1{-}x^2)(x^2{-}x^3)(x^3{-}x^1)$ is the basic anti-invariant, 
$j_q$ denotes the spherical Bessel function, and
\begin{equation}
\begin{aligned}
h_{\ell_2,\ell_3}^{(g,g')}(x)\ &\sim\
r^{6g +2g'+1+4\ell_2+6\ell_3}\,\De^{-g}\, \Xe^{-g'}\,
\tilde{\sigma}_2(\cD)^{\ell_2}\,\tilde{\sigma}_3(\cD)^{\ell_3}
\, \Xe^{g'}\,\De^g\,r^{-1-6g -2g'}\\
&\sim\ 
r^{6g +2g'+1+4\ell_2+6\ell_3}\,
\tilde{\sigma}_2(\widetilde{\cD})^{\ell_2}\,\tilde{\sigma}_3(\widetilde{\cD})^{\ell_3}
\,r^{-1-6g -2g'}
\end{aligned}
\end{equation}
is a deformed harmonic polynomial of degree $2\ell_2{+}3\ell_3$. 
Conjugation with $\Xe^{g'}\De^g$ defines the ``potential-free'' Dunkl operators
\begin{equation}
\widetilde{\cD}_i \= 
\partial_i\ +\sum_{\alpha\in R^+}\dfrac{g_{\alpha}\alpha_i}{(\alpha,x)}(1{-}s_\alpha)
\= \partial_i\ + \sum_{j(\neq i)}\frac{g}{x^i{-}x^j}(1{-}s_{i-j})
\ +\ \frac{g'}{x^1{+}x^2{+}x^3}(1{-}s_0) \ .
\end{equation}
The first polynomials read (up to a normalization constant)
\begin{equation}
\begin{aligned}
h^{(g,g')}_{0,0}(x) &\=1\ ,\\
h^{(g,g')}_{1,0}(x) &\=6(3g{+}1)\sigma_2-(6g{+}2g'{+}3)\tilde{\sigma}_2\ ,\\
h^{(g,g')}_{0,1}(x) &\=\tilde{\sigma}_3\ ,\\
h^{(g,g')}_{2,0}(x) &\=36(3g{+}1)(3g{+}2)\sigma_2^2 +(6g{+}2g'{+}5)(6g+
2g'{+}7)\tilde{\sigma}_2^2-12(3g{+}2)(6g{+}2g'{+}5)\sigma_2\tilde{\sigma}_2\ ,\\
h^{(g,g')}_{1,1}(x) &\= (6g{+}2g'{+}9)\tilde{\sigma}_2\tilde{\sigma}_3-6(3g{+}4)\sigma_2\tilde{\sigma}_3\ ,\\
h^{(g,g')}_{0,2}(x) &\=\kappa_1 \tilde{\sigma}_2^2 \sigma_2+\kappa_2 \tilde{\sigma}_2\sigma_2^2+\kappa_3 \sigma_2^3+\kappa_4 \tilde{\sigma}_3^2\ ,\\
h^{(g,g')}_{3,0}(x) &\=\kappa_5 \sigma_2^3 +\kappa_6 \tilde{\sigma}_2\sigma_2^2+\kappa_7\tilde{\sigma}_2^2 \sigma_2+\kappa_8 \tilde{\sigma}_2^3\ ,\\
h^{(g,g')}_{2,1}(x) &\=\kappa_9 \tilde{\sigma}_3 \sigma_2^2+\kappa_{10} \tilde{\sigma}_2\tilde{\sigma}_3\sigma_2+\kappa_{11}\tilde{\sigma}_2^2 \tilde{\sigma}_3\ ,
\end{aligned}
\end{equation}
where the coefficients $\kappa_i$ are given in Appendix~A.

\subsection{Intertwining Operators}
The $A_1{\oplus}A_2$ model features two independent anti-invariant polynomials,
\begin{equation}
\tau'_1(x) \= x^1{+}x^2{+}x^3 \= 3\,X \und
\tau_3(x) \= (x^1{-}x^2)(x^2{-}x^3)(x^3{-}x^1) \= \De\ ,
\end{equation} 
where the first one is invariant under $s_{i-j}$ and anti-invariant under $s_0$,
and the second one behaves oppositely.
They lead to the two intertwining operators
\begin{equation}\label{intA}
M'(g,g') \= \text{res}\bigl( \mathcal{D}_1{+}\mathcal{D}_2{+}\mathcal{D}_3 \bigr) \und
M(g,g') \= \text{res}\bigl( 
(\mathcal{D}_1{-}\mathcal{D}_2)(\mathcal{D}_2{-}\mathcal{D}_3)(\mathcal{D}_3{-}\mathcal{D}_1)\bigr)
\end{equation}
satisfying (\ref{Intrel}) for one of the couplings but not shifting the other one.
For the first intertwiner one easily finds
$M'(g,g')=\sum_i\partial_i-g'/X$ independent of~$g$. 
The second intertwiner is a more complicated expression;
the explicit form of $M(g,g'{=}0)$ is given in~\cite{CLP}.
The operators $\mathcal{R}(I)$ (\ref{R(g)}) associated to $M$ are obtained from the free case ($g{=}g'{=}0$),
\begin{equation}
\begin{aligned}
\mathcal{R}(I)&\=M(0,0)^2\=(\partial_1{-}\partial_2)^2(\partial_2{-}\partial_3)^2(\partial_3{-}\partial_1)^2 \\[4pt]
&\= -\tfrac{1}{27}\bigl((\pa_1{+}\pa_2{-}2\pa_3)(\pa_2{+}\pa_3{-}2\pa_1)(\pa_3{+}\pa_1{-}2\pa_2)\bigr)^2
+\tfrac{1}{54}\bigl((\pa_1{-}\pa_2)^2+(\pa_2{-}\pa_3)^2+(\pa_3{-}\pa_1)^2\bigr)^3 \\[4pt]
&\= \tfrac{1}{27}\tilde{I}_3^2-\tfrac{1}{54}\tilde{I}_2^3\ .
\end{aligned}
\end{equation}
For the $g'$ intertwiner~$M'$, we have
\begin{equation}
\mathcal{R'}(I)\=M'(0,0)^2\=(\pa_1{+}\pa_2{+}\pa_3)^2 
\= 3(\pa_1^2{+}\pa_2^2{+}\pa_3^2)-(\pa_1{-}\pa_2)^2-(\pa_2{-}\pa_3)^2-(\pa_3{-}\pa_1)^2
\=\tilde{I}_2{-} 3\,I_2\ .
\end{equation}

Applying a ($g{-}1$)-fold Darboux dressing with $M(h,0)$ for $h=1,2,\ldots g{-}1$ and $M(h,0)^*=M(-h,0)$ to
\begin{equation}
Q(1,0)\=M(0,0)\=(\partial_1{-}\partial_2)(\partial_2{-}\partial_3)(\partial_3{-}\partial_1)\ ,
\end{equation}
we obtain an exceptional independent conserved charge~$Q(g,0)$ 
in the case of integer coupling~$g$ and $g'{=}0$.
Likewise, Darboux dressing with $M'(0,h')$  and $M'(0,h')^*=M'(0,-h')$ of
\begin{equation}
Q'(0,1)\=M'(0,0)\=\partial_1{+}\partial_2{+}\partial_3
\end{equation}
produces such a charge~$Q'(0,g')$ for $g{=}0$ and integer~$g'$.
Combining both, by following a sequence of $M$ and $M'$ intertwiners 
starting either from $(1{-}g,g')$ to $(g,g')$ or from $(g,1{-}g')$ to $(g,g')$, 
we can extend these special charges to $Q(g,g')$ and $Q'(g,g')$, respectively, 
for all integral values of both couplings.
In such cases, the two extra charges 
\begin{equation}
\begin{aligned}
Q(g,g')&\=M(g{-}1,g')\,M(g{-}2,g')\cdots M(1,g')\,M(0,g')\,M(-1,g')\cdots M(2{-}g,g')\,M(1{-}g,g')\ ,\\[4pt]
Q'(g,g')&\=M'(g,g'{-}1)\,M'(g,g'{-}2)\cdots M'(g,1)\,M'(g,0)\,M'(g,-1)\cdots M'(g,2{-}g')\,M'(g,1{-}g')
\end{aligned}
\end{equation}
enhance the nonlinear algebra of integrals of motion to a $\Z_2\oplus\Z_2$ graded one, 
\begin{equation}
\begin{aligned}
&\im[\tilde{I}_2,\tilde{F}_2]\=2(3\tilde{I}_2 I_2-\tilde{I}_2^2)\ ,\quad 
\im[\tilde{I}_3,\tilde{F}_2]\=3(3\tilde{I}_3 I_2-\tilde{I}_2\tilde{I}_3)\ ,\\
&\im[\tilde{I}_2,\tilde{F}_3]\=2(3\tilde{I}_3 I_2-\tilde{I}_3\tilde{I}_2)\ ,\quad
\im[\tilde{I}_3,\tilde{F}_3]\=3(-\tilde{I}_3^2+\sfrac{3}{2}\tilde{I}_2^2 I_2)\ ,\\
&\im[\tilde{F}_2,\tilde{F}_3]\=\tfrac{1}{2}(\tilde{F}_3\tilde{I}_2+\tilde{I}_2\tilde{F}_3)-\tfrac{3}{2}(\tilde{F}_3 I_2+I_2\tilde{F}_3)\ ,\\
&\im[Q,\tilde{F}_2]\=-3(2g{-}1)Q( \tilde{I}_2-3 I_2)\ , \quad
\im[Q,\tilde{F}_3]\=-3(2g{-}1)Q \tilde{I}_3\ ,\\
&\im[Q',\tilde{F}_2]\= -(2g'{-}1)Q'\tilde{I}_2\ , \quad
\im[Q',\tilde{F}_3]\= -(2g'{-}1)Q' \tilde{I}_3\ ,\\
&Q^2\=(\mathcal{R}(I))^{2g-1}\ ,\quad
Q'^2\=(\mathcal{R'}(I))^{2g'-1}\ ,\quad
[Q,Q']\=0\ .
\end{aligned}
\end{equation}

\subsection{Translation-invariant limit}
It is instructive to consider the special case of $g'{=}0$, where translation invariance is recovered
and the center of mass feels no potential. 
In this limit, the reflection $s_0$ disappears from the consideration, 
and the total momentum appears as a first-order conserved charge.
It is then more convenient to replace~(\ref{polyA}) with the Newton sums
\begin{equation}
\begin{aligned}
\sigma_1(x) &\= x^1+x^2+x^3 \= 3\,X\ ,\\
\sigma_2(x) &\= (x^1)^2+(x^2)^2+(x^3)^2 \ ,\\
\sigma_3(x) &\= (x^1)^3+(x^2)^3+(x^3)^3 \ ,
\end{aligned}
\end{equation}
resulting in the basic Liouville integrals
\begin{equation}\label{intsmotionA0}
\begin{aligned}
I_1&\=-\im\,{\rm res}  \bigl(\mathcal{D}_1{+}\mathcal{D}_2{+}\mathcal{D}_3\bigr)\= P\ ,\\[6pt]
I_2&\=\ - {\rm res}  \bigl(\mathcal{D}_1^{2}{+}\mathcal{D}_2^{2}{+}\mathcal{D}_3^{2}\bigr) \=2\,H\ ,\\
I_3&\=\ \ \im\,{\rm res}  \bigl(\mathcal{D}_1^{3}{+}\mathcal{D}_2^{3}{+}\mathcal{D}_3^{3}\bigr)
\=\sum_{i} p_i^3+3\sum_{i<j}^3\frac{g(g{-}1)}{(x^i{-}x^j)^2}(p_i{+}p_j) \ .
\end{aligned}
\end{equation}
In this case, the two further integrals establishing superintegrability may be chosen as
\begin{equation}
F_1 \= \{H,3X\} - \half\{P,D\} \und F_3 \= \{H,J_3\} - \half\{I_3,D\}\ ,
\end{equation}
with
\begin{equation}
J_3 \= p_1x^1p_1{+}p_2x^2p_2{+}p_3x^3p_3\ +\ 
g(g{-}1) \bigl( \tfrac{x^1+x^2}{(x^1-x^2)^2}+\tfrac{x^2+x^3}{(x^2-x^3)^2}+\tfrac{x^3+x^1}{(x^3-x^1)^2}\bigr)\ .
\end{equation}

Due to the new first-order charge~$P$, the energy eigenfunctions carry a label $\ell_1$ rather than $\ell_2$,
\begin{equation}
\Psi^{(g)}_{\E,\ell_1,\ell_3}(x) \= 
j_q({\scriptstyle\sqrt{2E}}\,r)\,r^{-q}\De^g\, h^{(g)}_{\ell_1,\ell_3}(x) 
\qquad\textrm{with}\qquad q=3g+\ell_1+3\ell_3\ ,
\end{equation}
where now
\begin{equation}
\begin{aligned}
h_{\ell_1,\ell_3}^{(g)}(x)\ &\sim\
r^{6g+1+2\ell_1+6\ell_3}\,\De^{-g}\,\bigl(\cD_1{+}\cD_2{+}\cD_3\bigr)^{\ell_1}\,
\bigl(\cD_1^3{+}\cD_2^3{+}\cD_3^3\bigr)^{\ell_3}\,\De^g\,r^{-1-6g}\\
&\sim\ 
r^{6g+1+2\ell_1+6\ell_3}\,\bigl(\widetilde{\cD}_1{+}\widetilde{\cD}_2{+}\widetilde{\cD}_3\bigr)^{\ell_1}\,
\bigl(\widetilde{\cD}_1^3{+}\widetilde{\cD}_2^3{+}\widetilde{\cD}_3^3\bigr)^{\ell_3}\,r^{-1-6g}
\end{aligned}
\end{equation}
is a deformed harmonic polynomial of degree $\ell_1{+}3\ell_3$. The first polynomials can be written as
\begin{equation}
\begin{aligned}
h_{0,0}^{(g)}&=1\ , \\
h_{1,0}^{(g)}&=\sigma_1\ , \\ 
h_{2,0}^{(g)}&=(2g{+}1)\sigma_1^2{-}\sigma_2 \ , \\
h_{0,1}^{(g)}&=3(2g{+}1)\sigma_1\sigma_2{-}(6g{+}5)\sigma_3\ ,  \\
h_{3,0}^{(g)}&=-(6g{+}5)\sigma_1^3{+}9\sigma_2\sigma_1\ ,  \\
h_{1,1}^{(g)}&=-3 (2 g{+}1) (6 g{+}5) \sigma _2 \sigma _1^2{+}(6 g{+}5) (6 g{+}7) \sigma _3 \sigma _1{-}6 \sigma _2^2\ ,\\
h_{4,0}^{(g)}&=(6 g{+}5) (6 g{+}7) \sigma _1^4{-}18 (6 g{+}5) \sigma _2 \sigma _1^2{+}27 \sigma _2^2\ , \\
h_{2,1}^{(g)}&=(2 g{+}1) (6 g{+}7) \sigma _2 \sigma _1^3{-}(6 g{+}7) (2 g{+}3) \sigma _3 \sigma _1^2{-}3 (2 g{-}1) \sigma _2^2 \sigma _1{+}(6 g{+}7) \sigma _2 \sigma _3 \ ,\\
h_{5,0}^{(g)}&=-(6 g{+}7) (2 g{+}3) \sigma _1^5+10 (6 g{+}7) \sigma _2 \sigma _1^3-45 \sigma _2^2 \sigma _1\ , \\
h_{0,2}^{(g)}&=\gamma_1\sigma _2^2 \sigma _1^2{+}\gamma_2\sigma _2^3+\gamma_3 \sigma _2 \sigma _1^4+\gamma_4 \sigma _2 \sigma _3 \sigma _1+\gamma_5 \sigma _3^2\ , \\
h_{3,1}^{(g)}&=\gamma_6\sigma _2 \sigma _1^4+\gamma_7\sigma _3 \sigma _1^3+\gamma_8\sigma _2^2 \sigma _1^2+\gamma_9 \sigma _2 \sigma _3 \sigma _1+30 \sigma _2^3\ ,\\
h_{6,0}^{(g)}&=\gamma_{10} \sigma_1^6 + \gamma_{11}  \sigma_2\sigma_1^4 + \
\gamma_{12} \sigma_2^2\sigma_1^2  - 135 \sigma_2^3\  ,
\end{aligned}
\end{equation}
where the coefficients $\gamma_i$ are given in Appendix~A.
Roughly half of these polynomials (as linear combinations) can be obtained as the translation-invariant limit
$h_{\ell_2,\ell_3}^{(g,g'=0)}$ of the $A_1\oplus A_2$ polynomials, but new states arise in the limit, due to the
new first-order invariant $\sigma_1$.

Only the third-order intertwiner $M(g){\equiv}M(g,0)$ based on $\De$ remains, and via Darboux dressing it generates 
the extra charge~$Q(g){\equiv}Q(g,0)$ for integral values of~$g$, 
which enlarges the nonlinear algebra spanned by
$\{I_1,I_2,I_3,F_1,F_3\}$ to
\begin{equation}
\begin{aligned}
\im[I_1,F_1]&\=3I_2-I_1^2\ ,\qquad\qquad
\im[I_3,F_1]\=-3I_3 I_1+ 3I_2^2\ ,\\ 
\im[I_1,F_3]&\=-I_3 I_1+ I_2^2\ ,\qquad\quad\!
\im[I_3,F_3]\=-3I_3^2+4I_3 I_2 I_1+\tfrac32 I_2^3-3I_2^2 I_1^2+\tfrac12 I_2 I_1^4\ ,\\
\im[F_1,F_3]&\=\sfrac12(F_1I_3+I_3F_1+F_3 I_1+I_1 F_3)\ ,\\
\im[Q,F_1]&\=-3(2g{-}1)\,Q\,I_1\ , \quad\
\im[Q,F_3]\=-3(2g{-}1)\,Q\bigl(I_3-\tfrac{2}{3}I_2 I_1\bigr)\ ,\qquad
Q^2\=(\mathcal{R}(I))^{2g-1}\ .
\end{aligned}
\end{equation}

\newpage

\section{The $\bf{AD_3}$ model}
\subsection{Integrals of motion}
This rank-3 system is irreducible and simply-laced, so contains just a single coupling~$g$.
Depending on the choice of variables, it takes the $A_3$ or the $D_3$ form. Here, we choose the latter.
The Hamiltonian reads
\begin{equation}
H\=\tfrac{1}{2}\sum_i p_i^2\ + \sum_{i< j} \Big(\frac{g(g{-}1)}{(x^i{-}x^j)^2}+ \frac{g(g{-}1)}{(x^i{+}x^j)^2}\Big)\ .
\end{equation}
One can choose
\begin{equation}
\mathcal{R}_+=\{e_1{+}e_2,\hspace{0.2cm} e_1{+}e_3,\hspace{0.2cm} e_2{+}e_3,\hspace{0.2cm} 
e_1{-}e_2,\hspace{0.2cm} e_1{-}e_3, \hspace{0.2cm} e_2{-}e_3\}\ ,
\end{equation} 
leading to the Dunkl operators
\begin{equation}
\mathcal{D}_i\=\partial_i\ -\sum_{j(\neq i)} \Big(\frac{g}{x^i{-}x^j}s_{i-j}+\frac{g}{x^i{+}x^j}s_{i+j}\Big)
\end{equation}
with reflections $s_{i-j}$ given in (\ref{si-j}) and
\begin{equation} \label{si+j}
\begin{aligned}
s_{1+2}&: \ (x^1,x^2,x^3)\ \mapsto\ (-x^2,-x^1,x^3)\ , \\
s_{1+3}&: \ (x^1,x^2,x^3)\ \mapsto\ (-x^3,x^2,-x^1)\ , \\
s_{2+3}&: \ (x^1,x^2,x^3)\ \mapsto\ (x^1,-x^3,-x^2)\ . \\
\end{aligned}
\end{equation}
The lowest three Weyl-invariant polynomials are
\begin{equation}
\sigma_2(x) \= (x^1)^2+(x^2)^2+(x^3)^2 \ ,\qquad
\sigma_3(x) \= x^1\,x^2\,x^3\ ,\qquad
\sigma_4(x) \= (x^1)^4+(x^2)^4+(x^3)^4\ .
\end{equation}

Hence, the corresponding Liouville integrals read
\begin{equation}\label{intsmotionD}
\begin{aligned}
I_2&\=- {\rm res} \bigl(\mathcal{D}_1^{2}{+}\mathcal{D}_2^{2}{+}\mathcal{D}_3^{2}\bigr) \=2\,H\ ,\\
I_3&\= \ \im\,{\rm res} \bigl(\mathcal{D}_1\,\mathcal{D}_2\,\mathcal{D}_3 \bigr) \= 
p_1\,p_2\,p_3 - 4g(g{-}1)\Bigl( 
\tfrac{x^1x^2\,p_3}{((x^1)^2-(x^2)^2)^2} + 
\tfrac{x^2x^3\,p_1}{((x^2)^2-(x^3)^2)^2} + 
\tfrac{x^3x^1\,p_2}{((x^3)^2-(x^1)^2)^2} \Bigr)\ ,\\
I_4&\= \ \ {\rm res} \bigl(\mathcal{D}_1^{4}{+}\mathcal{D}_2^{4}{+}\mathcal{D}_3^{4}\bigl) \= p_1^4+2g(g{-}1)\textstyle \sum_{\ell\neq 1} \left\{ p_1^2,  \sfrac{1}{(x^1-x^\ell)^2}{+}\sfrac{1}{(x^1+x^\ell)^2}\right\}+16g(g{-}1) \sfrac{x^1 x^2}{((x^1)^2-(x^2)^2)^2}p_1p_2\\ 
&-2 \im g(g{-}1)\textstyle \sum_{\ell\neq 1}\left\{  p_1,  \sfrac{1}{(x^1-x^\ell)^3}{+}\sfrac{1}{(x^1+x^\ell)^3}\right\} +16 g^2(g{-}1)^2\left( \sfrac{(x^1)^4+(x^2)^4}{((x^1)^2-(x^2)^2)^4}+\sfrac{(x^1)^2+(x^2)^2}{((x^1)^2-(x^2)^2)^2}\sfrac{(x^1)^2+(x^3)^2}{((x^1)^2-(x^3)^2)^2} \right)\\
&+\text{cyclic}\ ,
\end{aligned}
\end{equation}
where the term
``cyclic'' refers to adding the cyclic permutations of the labels (1,2,3).
The two lowest additional conserved charges are
\begin{equation}
F_3 \= \{H,J_3\}-\half\{I_3,D\} \und
F_4 \= \{H,J_4\}-\half\{I_4,D\} \ ,
\end{equation}
with 
\begin{equation}
\begin{aligned}
J_3 &\= \tfrac{1}{3}(x^1p_2p_3{+}x^2p_3p_1{+}x^3p_1p_2)-\sfrac{4}{3}g(g{-}1)\Bigl(\tfrac{1}{(x^1-x^2)^2(x^1+x^2)^2}{+}\tfrac{1}{(x^2-x^3)^2(x^2+x^3)^2}{+}\tfrac{1}{(x^3-x^1)^2(x^3+x^1)^2}\Bigr)x^1 x^2 x^3\ ,\\
J_4 &\=\sfrac{1}{2}\{x^1,p_1^3 \}+4g(g{-}1)\left(\sfrac{(x^1)^2+2(x^2)^2}{((x^1)^2-(x^2)^2)^2}+\sfrac{(x^1)^2+2(x^3)^2}{((x^1)^2-(x^3)^2)^2}\right) x^1 p_1 -2\mi g(g{-}1)\sfrac{(x^1)^2+(x^3)^2}{((x^1)^2-(x^3)^2)^2}+\text{cyclic}\ .\end{aligned}
\end{equation}

\newpage

\subsection{Energy eigenstates}
For the eigenvalue problem
\begin{align}
H \, \Psi^{(g)}_{\E,\ell_3,\ell_4}(x)  \= E\,  \Psi^{(g)}_{\E,\ell_3,\ell_4}(x) 
\end{align}
one obtains the energy eigenfunctions
\begin{equation}
\Psi^{(g)}_{\E,\ell_3,\ell_4}(x) \ \equiv\ \<x \mid \ell_3,\ell_4\>_g \= 
j_q({\scriptstyle\sqrt{2E}}\,r)\,r^{-q}\De^g\,
h^{(g)}_{\ell_3,\ell_4}(x) 
\qquad\textrm{with}\qquad q=6g+3\ell_3+4\ell_4\ ,
\end{equation}
where  $\De=(x^1{-}x^2)(x^1{+}x^2)(x^2{-}x^3)(x^2{+}x^3)(x^3{-}x^1)(x^3{+}x^1)$
is the basic anti-invariant, and
\begin{equation}
\begin{aligned}
h^{(g)}_{\ell_3,\ell_4}(x) \ &\sim\
r^{12g+1+6\ell_3+8\ell_4}\,\Delta^{-g}\,\bigl(\cD_1\cD_2\cD_3\bigr)^{\ell_3}\,
\bigl(\cD_1^4{+}\cD_2^4{+}\cD_3^4\bigr)^{\ell_4}\,\De^g\,r^{-1-12g} \\
&\sim\ 
r^{12g+1+6\ell_3+8\ell_4}\,\bigl(\widetilde{\cD}_1\widetilde{\cD}_2\widetilde{\cD}_3\bigr)^{\ell_3}\,
\bigl(\widetilde{\cD}_1^4{+}\widetilde{\cD}_2^4{+}\widetilde{\cD}_3^4\bigr)^{\ell_4}\,r^{-1-12g}
\end{aligned}
\end{equation}
is a deformed harmonic polynomial of degree $3\ell_3{+}4\ell_4$. 
Conjugation with $\De^g$ defines the ``potential-free'' Dunkl operators
\begin{equation}
\widetilde{\cD}_i \= \
\partial_i\ +\sum_{j(\neq i)} \Big(\frac{g}{x^i{-}x^j}(1{-}s_{i-j})+\frac{g}{x^i{+}x^j}(1{-}s_{i+j})\Big)\ .
\end{equation}
The first polynomials read
	\begin{equation}
	\label{ADstates}
	\begin{aligned}
	h^{(g)}_{0,0}(x) &\=1\ ,\\
	h^{(g)}_{1,0}(x) &\=\sigma_3\ ,\\
	h^{(g)}_{0,1}(x) &\=(12g{+}5) \sigma_4 - (8 g{+}3) \sigma_2^2\ ,\\
	h^{(g)}_{2,0}(x) &\=\alpha_1 \sigma_2^3 + \alpha_2 \sigma_3^2 + 
	\alpha_3 \sigma_2 \sigma_4\ ,\\
	h^{(g)}_{1,1}(x) &\=\alpha_4 \sigma_2^2 \sigma_3 +\alpha_5 \sigma_4  \sigma_3\ ,\\
	h^{(g)}_{0,2}(x) &\=\alpha_6 \sigma_2^4 +\alpha_7\sigma_2 \sigma_3^2 +\alpha_8\sigma_2^2 \sigma_4 + 
\alpha_9 \sigma_4^2\ ,\\
	h^{(g)}_{3,0}(x) &\=\alpha_{10}\sigma_3^3 +\alpha_{11} \sigma_2  \sigma_3 \sigma_4 + \alpha_{12}\sigma_2^3 \sigma_3\ ,\\
	h^{(g)}_{2,1}(x) &\=\alpha_{13} \sigma_2^5 +\alpha_{14} \sigma_2^2 \sigma_3^2 +\alpha_{15}\sigma_2^3 \sigma_4 + 
	\alpha_{16} \sigma_3^2 \sigma_4 + 
\alpha_{17} \sigma_2 \sigma_4^2\ ,\\	
	h^{(g)}_{1,2}(x) &\=\alpha_{18} \sigma_2^4 \sigma_3 + 
	\alpha_{19} \sigma_2 \sigma_3^3 + 
	\alpha_{20}\sigma_2^2 \sigma_3 \sigma_4 +\alpha_{21} \sigma_3 \sigma_4^2\ ,\\
        h^{(g)}_{0,3}(x) &\=\alpha_{22} \sigma_2^6 + \alpha_{23}\sigma_2^3 \sigma_3^2 + \alpha_{24} \sigma_2^4 \sigma_4 + \alpha_{25}\sigma_2 \sigma_3^2 \sigma_4 + \alpha_{26}\sigma_2^2 \sigma_4^2 + \alpha_{27} \sigma_4^3  \ ,\\
        h^{(g)}_{4,0}(x) &\= \alpha_{28} \sigma_2^6 + \alpha_{29}\sigma_2^3 \sigma_3^2 + \alpha_{30} \sigma_3^4 + \alpha_{31}\sigma_2^4 \sigma_4 + \alpha_{32}\sigma_2 \sigma_3^2 \sigma_4 + \alpha_{33} \sigma_2^2 \sigma_4^2 \ ,
	\end{aligned}
	\end{equation}
where the coefficients $\alpha_i$ are given in Appendix~A.
\subsection{Intertwining Operators}
The basic anti-invariant of the $AD_3$ model reads
\begin{equation}
\tau_6(x) \= (x^1{-}x^2)(x^1{+}x^2)(x^2{-}x^3)(x^2{+}x^3)(x^3{-}x^1)(x^3{+}x^1) \= \De\ ,
\end{equation}
which produces the intertwiner
\begin{equation}\label{inD}
M(g)\=\text{res}\bigl( 
(\mathcal{D}_1^2{-}\mathcal{D}_2^2)
(\mathcal{D}_2^2{-}\mathcal{D}_3^2)
(\mathcal{D}_3^2{-}\mathcal{D}_1^2)  \bigr)\ ,
\end{equation}
satisfying (\ref{Intrel}).
The polynomial $\mathcal{R}(I)$ can be computed from the free case ($g{=}0$),
\begin{equation}
\begin{aligned}
\mathcal{R}(I)&\= M(0)^2 \=
(\partial_1^2{-}\partial_2^2)^2(\partial_2^2{-}\partial_3^2)^2(\partial_3^2{-}\partial_1^2)^2 \\
&\=\tfrac12 I_4^3-\tfrac54 I_4^2 I_2^2-9 I_4 I_3^2 I_2+ I_4 I_2^4-27 I_3^4+5 I_3^2 I_2^3-\tfrac14 I_2^6\ .
\end{aligned}
\end{equation}

When the coupling $g$ takes an integral value, there exists a sixth independent conserved charge
\begin{equation}
Q(g)\=M(g{-}1)\,M(g{-}2)\cdots M(1)\,M(0)\,M(-1)\cdots M(2{-}g)\,M(1{-}g)\ ,
\end{equation}
which extends the nonlinear algebra spanned by $\{I_2,I_3,I_4,F_3,F_4\}$ to
\begin{equation}
\begin{aligned}
\mi[I_3,F_3] &\= \tfrac16(I_2^3{-}I_4 I_2{-}18I_3^2)  \ ,\qquad
\mi[I_4,F_3] \= \tfrac43 (I_3 I_2^2{-}3 I_4 I_3)  \ ,\\
\mi[I_3,F_4] &\=  I_3 I_2^2{-}3 I_4 I_3 \ ,\qquad\qquad\ \
\mi[I_4,F_4] \= 2({-} I_2^4{+}3I_4 I_2^2{+}6 I_3^2 I_2 {-}2 I_4^2 )\ ,\\
\mi[F_3,F_4] &\=\{F_4,I_3\}  -\tfrac{1}{2}\{F_3,I_4{-}I_2^2\}\ ,\\
\mi[Q,F_3] &\=-6 (2g{-}1)\,Q\,I_3\ ,\quad
\mi[Q,F_4] \=6 (2g{-}1)\,Q\,(\tfrac23 I_2^2-I_4)\ ,\quad
Q^2 \= (\mathcal{R}(I))^{2g-1}\ .
\end{aligned}
\end{equation}

\section{The $\bf{BC_3}$ model}
\subsection{Integrals of motion}
This is the only irreducible non-simply-laced rank-3 model, so we have to deal with two coupling constants, $\gl$ and~$\gs$. 
It is described by the Hamiltonian
\begin{equation}\label{HB3}
H\=\tfrac{1}{2}\sum_i^3 p_i^2\ +
\sum_{i<j}\Big(\frac{\gl(\gl{-}1)}{(x^i{-}x^j)^2}+\frac{\gl(\gl{-}1)}{(x^i{+}x^j)^2}\Big)\ +\sum_i\frac{\gs(\gs{-}1)}{2\,(x^i)^2}\ .
\end{equation}
We take the set of positive roots as
\begin{equation}
\mathcal{R}_+\=\{e_1,\hspace{0.2cm} e_2,\hspace{0.2cm} e_3,\hspace{0.2cm} 
e_1{+}e_2,\hspace{0.2cm} e_1{+}e_3,\hspace{0.2cm} e_2{+}e_3,\hspace{0.2cm} e_1{-}e_2,\hspace{0.2cm} e_1{-}e_3, \hspace{0.2cm} e_2{-}e_3\}\ ,
\end{equation}
so the Dunkl operators read
\begin{equation}
\mathcal{D}_i\=\partial_i\ -\sum_{j(\neq i)} \Big(\frac{\gl}{x^i{-}x^j}s_{i-j}+\frac{\gl}{x^i{+}x^j}s_{i+j}\Big)\ -\ \frac{\gs}{x^i}s_{i}\ ,
\end{equation}
where the reflections are given in (\ref{si-j}) and (\ref{si+j}) and by
\begin{equation}
\begin{aligned}
s_1 &: \ (x^1,x^2,x^3)\ \mapsto\ (-x^1,x^2,x^3)\ ,\\
s_2 &: \ (x^1,x^2,x^3)\ \mapsto\ (x^1,-x^2,x^3)\ ,\\
s_3 &: \ (x^1,x^2,x^3)\ \mapsto\ (x^1,x^2,-x^3)\ .
\end{aligned}
\end{equation}
The three lowest Weyl-invariant polynomials may be chosen as
\begin{equation} \label{sigmaB3}
\sigma_2(x) = (x^1)^2+(x^2)^2+(x^3)^2 \ ,\qquad
\sigma_4(x) = (x^1)^4+(x^2)^4+(x^3)^4\ ,\qquad
\sigma_6(x) = (x^1\,x^2\,x^3)^2 = \sigma_3^2\ ,
\end{equation}
but one may also remain with the even Newton sums, replacing $\sigma_6$ with
$\sigma'_6(x)=(x^1)^6+(x^2)^6+(x^3)^6$.

For the basis~(\ref{sigmaB3}),
the first three Liouville integrals take the form
\begin{equation}\label{intsmotionB}
\begin{aligned}
I_2&\=- {\rm res} \bigl(\mathcal{D}_1^{2}{+}\mathcal{D}_2^{2}{+}\mathcal{D}_3^{2}\bigr) \=2\,H\ ,\\
I_4&\= \ \ {\rm res} \bigl(\mathcal{D}_1^{4}{+}\mathcal{D}_2^{4}{+}\mathcal{D}_3^{4}\bigl) \ ,\\
I_6&\=- {\rm res} \bigl(\mathcal{D}_1^2\,\mathcal{D}_2^2\,\mathcal{D}_3^2 \bigr) \ ,
\end{aligned}
\end{equation}
with the explicit form of $I_4$ and $I_6$ displayed in Appendix~B.
Two additional integrals of motion (not in involution) are
\begin{equation}
F_4 \=  \{H,J_4\}-\half\{I_4,D\}\und
F_6 \=  \{H,J_6\}-\half\{I_6,D\}\ ,
\end{equation}
where $J_4$ and $J_6$ are also given in Appendix~B.

\subsection{Energy eigenstates}
The eigenvalue problem
\begin{align}
H \, \Psi^{(\gl,\gs)}_{\E,\ell_4,\ell_6}(x) \= E\, \Psi^{(\gl,\gs)}_{\E,\ell_4,\ell_6}(x) 
\end{align}
is solved by
\begin{equation}
\Psi^{(\gl,\gs)}_{\E,\ell_4,\ell_6}(x) \ \equiv\ \<x \mid \ell_4,\ell_6\>_{g,g'}  
\= j_q({\scriptstyle\sqrt{2E}}\,r)\,r^{-q}\Delta_\ell^{\gl}\Delta_s^{\gs}\,
h^{(\gl,\gs)}_{\ell_4,\ell_6}(x) 
\qquad\textrm{with}\qquad q=6\gl+3\gs+4\ell_4+6\ell_6\ ,
\end{equation}
where $\Delta_\ell=(x^1{-}x^2)(x^1{+}x^2)(x^2{-}x^3)(x^2{+}x^3)(x^3{-}x^1)(x^3{+}x^1)$ and
$\Delta_s=x^1\,x^2\,x^3$, as well as
\begin{equation}
\begin{aligned}
h^{(\gl,\gs)}_{\ell_4,\ell_6}(x) \ &\sim\
r^{12\gl+6\gs+1+8\ell_4+12\ell_6}\,\Delta_\ell^{-\gl}\Delta_s^{-\gs}\,
\bigl(\cD_1^4{+}\cD_2^4{+}\cD_3^4\bigr)^{\ell_4}\,
\bigl(\cD_1\cD_2\cD_3\bigr)^{2\ell_6}\,
\De^g\,r^{-1-12\gl-6\gs} \\
&\sim\ 
r^{12\gl+6\gs+1+8\ell_4+12\ell_6}\,
\bigl(\widetilde{\cD}_1^4{+}\widetilde{\cD}_2^4{+}\widetilde{\cD}_3^4\bigr)^{\ell_4}\,
\bigl(\widetilde{\cD}_1\widetilde{\cD}_2\widetilde{\cD}_3\bigr)^{2\ell_6}\,r^{-1-12\gl-6\gs}
\end{aligned}
\end{equation}
being a deformed harmonic polynomial of degree $4\ell_4{+}6\ell_6$. 
The ``potential-free'' Dunkl operators read
\begin{equation}
\widetilde{\cD}_i \= \
\partial_i\ +\sum_{j(\neq i)} \Big(\frac{\gl}{x^i{-}x^j}(1{-}s_{i-j})+\frac{\gl}{x^i{+}x^j}(1{-}s_{i+j})\Big)\ +\  \frac{\gs}{x^i}(1{-}s_{i})\ .
\end{equation}
The first  polynomials read
	\begin{equation}
	\begin{aligned}
	h^{(\gl,\gs)}_{0,0}(x) &\=1\ ,\\
	h^{(\gl,\gs)}_{1,0}(x) &\=-(8 \gl {+} 2 \gs{+}3) \sigma_2^2 + (12 \gl {+} 6 \gs{+}5) \sigma_4\ ,\\
	h^{(\gl,\gs)}_{0,1}(x) &\= \mu_1 \sigma_2^3 + 
\mu_2 \sigma_2 \sigma_4 + \mu_3 \sigma_6\ ,\\
	h^{(\gl,\gs)}_{2,0}(x) &\=\mu_4 \sigma_2^4 +\mu_5 \sigma_2^2 \sigma_4 + 
\mu_6 \sigma_4^2 +\mu_7 \sigma_2 \sigma_6\ ,\\
	h^{(\gl,\gs)}_{1,1}(x) &\=\mu_8 \sigma_2^5 +\mu_9 \sigma_2^3 \sigma_4 + 
\mu_{10} \sigma_2 \sigma_4^2 +\mu_{11} \sigma_2^2 \sigma_6 + \mu_{12} \sigma_4 \sigma_6\ ,\\
	h^{(\gl,\gs)}_{0,2}(x) &\=\mu_{13} \sigma_2^6 +\mu_{14} \sigma_2^4 \sigma_4 + 
\mu_{15} \sigma_2^2 \sigma_4^2 +\mu_{16} \sigma_2^3 \sigma_6 +
\mu_{17} \sigma_2 \sigma_4 \sigma_6 +\mu_{18} \sigma_6^2\ ,\\
        h^{(\gl,\gs)}_{3,0}(x) &\=\mu_{19} \sigma_2^6 + \mu_{20} \sigma_2^4 \sigma_4 + \mu_{21} \sigma_2^2 \sigma_4^2  + \mu_{22} \sigma_2^3 \sigma_6 + \mu_{23} \sigma_2 \sigma_4 \sigma_6 + \mu_{24} \sigma_4^3 \ .
	\end{aligned}
	\end{equation}
where the coefficients $\mu_i$ are given in Appendix~A.
It is obvious that putting $\gs=0$ and $\gl=g$ brings us to the corresponding states~(\ref{ADstates}).

\subsection{Intertwining Operators}
The short-root and long-root anti-invariant polynomials are
\begin{equation}
\tau'_3(x) = x^1\,x^2\,x^3 = \De_s \und 
\tau_6(x) = (x^1{-}x^2)(x^1{+}x^2)(x^2{-}x^3)(x^2{+}x^3)(x^3{-}x^1)(x^3{+}x^1) = \De_\ell\ ,
\end{equation}
yielding the short-root and long-root intertwiners 
\begin{equation}\label{Mgs}
M_{s}(\gl,\gs)\=\text{res}\prod_{i}\mathcal{D}_i \und
M_{\ell}(\gl,\gs)\=\text{res}\prod_{i<j}(\mathcal{D}^2_i{-}\mathcal{D}^2_j)\ ,
\end{equation}
respectively.
They satisfy the relations (\ref{Intrel}), such that
\begin{equation}
\begin{aligned}
M_{s}(\gl,\gs)\,I_k(\gl,\gs)\=I_k(\gl,\gs{+}1)\,M_{s}(\gl,\gs) \ ,\\
M_{\ell}(\gl,\gs)\,I_k(\gl,\gs)\=I_k(\gl{+}1,\gs)\,M_{\ell}(\gl,\gs)\ .
\end{aligned}
\end{equation}
From the free case ($\gl{=}\gs{=}0$), one can compute the operators $\mathcal{R}_{s}(I)$ and $\mathcal{R}_{\ell}(I)$ (\ref{R(g)}),
\begin{equation}
\begin{aligned}
\mathcal{R}_{s}(I)&\=\prod_{i}\partial_i^2 \qquad\quad\!\= - I_6\ , \\
\mathcal{R}_{\ell}(I)&\=\prod_{i<j}(\partial^2_i{-}\partial^2_j)^2 \= 
-27 I_6^2 -9 I_6 I_4 I_2 + 5 I_6 I_2^3 + \tfrac12 I_4^3 -\tfrac54 I_4^2 I_2^2 + I_4 I_2^4 -\tfrac14 I_2^6\ .
\end{aligned}
\end{equation}

Finally, for integral values of both couplings we can construct two more conserved charges,
\begin{equation}
\begin{aligned}
Q_\ell(\gl,\gs) &\=  
M_\ell(\gl{-}1,\gs)\,\cdots\,M_\ell(1,\gs)\,M_\ell(0,\gs)\,M_\ell(-1,\gs)\,\cdots\,M_\ell(1{-}\gl,\gs)\ ,\\
Q_s(\gs,\gl) &\=  
M_s(\gl,\gs{-}1)\,\cdots\,M_s(\gl,1)\,M_s(\gl,0)\,M_s(\gl,-1)\,\cdots\,M_s(\gl,1{-}\gs)\ .
\end{aligned}
\end{equation}
Together with $F_k$ and $I_k$, they satisfy the following nonzero relations,
\begin{equation}
\begin{aligned}
\mi[I_4,F_4]&\= 2({-} I_2^4{+}3I_4 I_2^2{+}6 I_6 I_2 {-}2 I_4^2 ), \ ,\qquad
\mi[I_6,F_4]\= 2 (I_6 I_2^2{-}3 I_6 I_4)   \ ,\\
\mi[I_4,F_6]&\= \tfrac43 I_6 I_2^2{-}4 I_6 I_4  \ ,\qquad
\mi[I_6,F_6]\= \tfrac13(I_6 I_2^3{-}I_6 I_4 I_2{-}18I_6^2)  \ ,\\
\mi[F_4,F_6]&\=\{F_6,2I_4{-}I_2^2\}-\{F_4,I_6\}  \ ,\\
\mi[Q_{s},F_4]&\=-3(2\gs{-}1)Q_{s}(I_4{-}\sfrac{1}{3}I_2^2)\ ,\qquad
\mi[Q_{s},F_6]\=-3(2\gs{-}1)Q_{s}(I_6{+}\tfrac{1}{18}I_2I_4{-}\tfrac{1}{18}I_2^3)\ ,\\
\mi[Q_{\ell},F_4]&\=-6(2\gl{-}1)Q_{\ell}(I_4{-}\sfrac{2}{3}I_2^2)\ ,\qquad
\mi[Q_{\ell},F_6]\=-6(2\gl{-}1)Q_{\ell}I_6\ ,\\
Q_{s}^2&\=(\mathcal{R}_{s}(I))^{2\gs-1} \ ,\qquad
Q_{\ell}^2=(\mathcal{R}_{\ell}(I))^{2\gl-1} \ ,\qquad
[Q_{s},Q_{\ell}] \=0\ ,
\end{aligned}
\end{equation}
which define a $\Z_2{\oplus}\Z_2$ graded polynomial algebra of conserved charges.

\vskip 1cm
\noindent
{\bf Acknowledgments:} 
FC-M was partially supported by Fondecyt grant 1171475. 
FC was partially supported by Fondecyt grant 1171475 and Becas Santander Iberoam\'erica. FC would like to thank the Departamento de F\'isica Te\'orica, \'Atomica y \'Optica at the Universidad de Valladolid for all the support and kind hospitality. OL ist grateful to the Instituto de Ciencias F\'isicas y Matem\'aticas at the Universidad Austral de Chile for warm hospitality.

\vskip 2cm

\appendix

\section{Polynomials coefficients}

\subsubsection*{{\bf $A_1\oplus A_2$} model}
\vspace{-8mm}
\begin{multicols}{2}
\begin{equation*}
\begin{aligned}
\kappa_1&=-27(6g{+}2g'{+}7)(6g{+}2g'{+}9)\ , \\
\kappa_2&=162(3g{+}2)(6g{+}2g'{+}7)\ ,\\
\kappa_3&=-324(3g{+}1)(3g{+}2)\ , \\
\kappa_4&=2(6g{+}2g'{+}7)(6g{+}2g'{+}9)(6g{+}2g'{+}11)\ ,\\
\kappa_5&=-648(g{+}1)(3g{+}1)(3g{+}2)\ ,\\
\kappa_6&=324(g{+}1)(3g{+}2)(6g{+}2g'{+}7)\ ,
\end{aligned}
\end{equation*}
\begin{equation}
\begin{aligned}
\\
\kappa_7&=-54(g{+}1)(6g{+}2g'{+}7)(6g{+}2g'{+}9)\ , \\
\kappa_8&=(6g{+}2g'{+}7)(6g{+}2g'{+}9)(6g{+}2g'{+}11)\ ,\\
\kappa_9&=-36(3g{+}4)(3g{+}5)\ ,\\
\kappa_{10}&=12(3g{+}5)(6g{+}2g'{+}11)\ ,\\
\kappa_{11}&=-(6g{+}2g'{+}11)(6g{+}2g'{+}13)\ .
\end{aligned}
\end{equation}
\end{multicols}

\subsubsection*{{\bf $A_2$} model }
\vspace{-8mm}
\begin{multicols}{2}
\begin{equation*}
\begin{aligned}
\gamma_1&=6 (6 g{+}7) \left(4 g^2{+}11 g{+}10\right)\ ,\\ 
\gamma_2&=-3 \left(4 g^2{+}14 g{+}17\right)\ , \\ 
\gamma_3&={-}3 (2 g{+}3) (6 g{+}7)\ ,  \\
\gamma_4&={-}12 (2 g{+}3)^2 (6 g{+}7)\ ,\\
\gamma_5&=2 (2 g{+}3) (6 g{+}7) (6 g{+}11)\ ,\\
\gamma_6&=-3 (2 g{+}1) (2 g{+}3) (6 g{+}7)\ ,
\end{aligned}
\end{equation*}
\begin{equation}
\begin{aligned}
\\
\gamma_7&=(2 g{+}3) (6 g{+}7) (6 g{+}11)\ ,\\
\gamma_8&=3 (6 g{-}1) (6 g{+}7)\ ,\\
\gamma_9&= -9 (2 g{+}3) (6 g{+}7)\ ,\\
\gamma_{10}&=(2g{+}3) (6 g{+}7) (6 g{+}11) \ ,\\
\gamma_{11}&=-45(2 g{+}3) (6 g{+}7) \ ,\\
\gamma_{12}&=135(6 g{+}7) \ .\\
\end{aligned}
\end{equation}
\end{multicols}

\newpage

\subsubsection*{{\bf $AD_3$} model }
\vspace{-8mm}
\begin{multicols}{2}
		\begin{equation*}
		\begin{aligned}
		\alpha_1&={-}(28 g{+}17)\ ,\\
		\alpha_2&=6 (12 g{+}7) (12 g{+}11)\ ,\\
		\alpha_3&=3 (12 g{+}7)\ ,\\
		\alpha_4&=(8 g{+}5)\ ,\\
		\alpha_5&=- (12 g{+}11)\ ,\\
		\alpha_6&=(64 g^2{+}152 g {+}99 )\ ,\\
		\alpha_7&=- 48 (12 g{+}13 ) \ ,\\
		\alpha_8&=- 
		6 (32 g^2{+} 76 g{+}47  ) \ ,\\
		\alpha_9&=	3 (4 g{+} 5) (12 g {+}13)\ ,\\
		\alpha_{10}&=-2(12 g{+} 13) (12 g{+}17)\ ,\\
		\alpha_{11}&=- 
		3 (12 g{+}13)\ ,\\
		\alpha_{12}&=(28 g{+}27)\ ,\\
		\alpha_{13}&=(224 g^2 {+} 492 g{+}255)\ ,\\
		\alpha_{14}&=- 
		6 (4 g{+}5) ( 12 g{+}11) (24 g{+}29)\ ,\\
		\alpha_{15}&=- 
		4 (12 g{+}11) (13 g{+}18)\ , \\
		\alpha_{16}&=6 (12 g{+}11) (12 g{+}17) (12 g{+}19 )\ ,\\
		\alpha_{17}&=	3 (12 g{+}11) (12 g{+}17 )\ , \\
		\alpha_{18}&=-(64 g^2{+} 184 g{+}159 )\ ,
		\end{aligned}
		\end{equation*}
		\begin{equation}
		\begin{aligned}
		\\
		\alpha_{19}&=48 (12 g{+}19)\ ,\\
		\alpha_{20}&=6 (32 g^2{+}100 g{+}83)\ , \\
		\alpha_{21}&=-3 (4 g{+}7 ) (12 g{+}19)\ ,\\
		\alpha_{22}&=-(73728 g^5{+}589824 g^4{+}1898752g^3{+}3045328g^2{+}2416240g\\
		&{+}754803)\ ,\\
		\alpha_{23}&=48 (12 g{+}13)(12 g{+}17) (288 g^2{+}868 g{+}651)\ ,\\
		\alpha_{24}&=3 (12 g{+}13)(9216 g^4{+}63360 g^3{+}164912 g^2  {+}191336 g {+} 83097)\ ,\\
		\alpha_{25}&=-432 (4 g{+}7)(12 g{+} 13) (12 g{+}17) (12g{+}19)\ ,\\
		\alpha_{26}&=-3 (12 g{+}13)(12 g{+}17) (12 g{+}19) (96 g^2{+}364g{+}357) \ ,\\
		\alpha_{27}&=3 (4 g{+}7) (12g{+}13) (12 g{+}17) (12 g{+}19) (12 g{+}23)\ ,\\
		\alpha_{28}&=(3136 g^3{+}12656 g^2 {+}16764g{+} 7269)\ ,\\
		\alpha_{29}&=-12 (12 g{+}13) (12 g{+} 17) (112 g^2{+}344g{+}263)\ ,\\
		\alpha_{30}&=12 (4g{+}7) (12 g{+}13) (12 g{+}17) (12 g{+}19) (12 g{+} 23)\ ,\\
		\alpha_{31}&=-6 (12 g{+} 13) (112 g^2{+} 336 g {+} 251 )\ ,\\
		\alpha_{32}&=36 (4 g{+}7) (12 g{+}13) (12 g{+}17) (12 g{+}19)\ ,\\
		\alpha_{33}&=3 (12 g{+}13) (12 g{+}17) (12 g{+}19)\ .
		\end{aligned}
		\end{equation}
\end{multicols}

\subsubsection*{{\bf $BC_3$} model }
\vspace{-8mm}
\begin{multicols}{2}
		\begin{equation*}
		\begin{aligned}
		\mu_1&=-(2 \gs{+}1) (28 \gl + 10 \gs{+}17)\ , \\
		\mu_2&=3 (2 \gs{+}1) (12 \gl + 6 \gs{+}7)\ ,\\
		\mu_3&=6 ( 12 \gl {+} 6 \gs{+}7) (12 \gl {+} 6 \gs{+}11)\ ,\\
		\mu_4&=(64 \gl^2{+} 4 \gs^2 {+} 56 \gs {+} 8 \gl (4 \gs{+}19){+}99)\ ,\\
		\mu_5&=- 
		6 (32 \gl^2{+} 4 \gs^2 {+} 32 \gs {+} 
		4 \gl (6 \gs{+}19 ){+} 47 )\ ,\\
		\mu_6&=3 (4 \gl {+} 2 \gs{+}5 ) ( 12 \gl {+} 6 \gs {+}13) \ ,\\
		\mu_7&=- 
		48 (12 \gl {+} 6 \gs{+}13)\ , \\
		\mu_8&=( 2 \gs{+}1 ) ( 224 \gl^2 {+} 20 \gs^2 {+} 168 \gs {+} 4 \gl (34 \gs{+}123){+}255 )\ , \\
		\mu_9&=-4 ( 2 \gs{+}1) ( 13 \gl {+} 4 \gs{+}18) ( 12 \gl {+} 6 \gs{+}11 )\ , \\
		\mu_{10}&=3 (2 \gs{+}1) ( 12 \gl {+} 6 \gs{+}11) ( 12 \gl {+} 6 \gs{+}17)\ ,\\
		\mu_{11}&=- 
		6 (12 \gl {+} 6 \gs{+}11) (96 \gl^2{+} 12 \gs^2 {+} 104 \gs {+} 
		4 \gl ( 18 \gs{+}59){+}145)\ ,\\
		\mu_{12}&=	6 ( 12 \gl {+} 6 \gs{+}11) (12 \gl {+} 6 \gs{+}17) (12 \gl {+} 6 \gs{+}19)\ , \\
			\mu_{13}&=(2\gs{+} 1)(2\gs{+}3) (4 \gl (4191{+}3164 \gl{+}784 \gl^2){+}16 \gl \gs (645{+}238 \gl{+}95 \gs){+}2 \gs (3447{+}1046 \gs{+}100 \gs^2){+}7269)\ ,\\
		\mu_{14}&=-6 (2 \gs{+}1) ( 2 \gs{+} 3) ( 12 \gl {+} 6 \gs{+}13 ) (112 \gl ( \gl{+}3 ) {+} 
		144 \gs {+} 96 \gl \gs {+} 20 \gs^2{+}251  )\ ,\\
		\mu_{15}&=3 ( 2 \gs{+}1 ) ( 2 \gs {+}3) ( 12 \gl {+} 6 \gs {+}13) ( 12 \gl {+} 
		6 \gs{+} 17) ( 12 \gl {+} 6 \gs{+}19 )\ ,\\		
		\mu_{16}&=- 
		12 (2 \gs{+}3  ) ( 12 \gl {+} 6 \gs {+}13) ( 12 \gl {+} 6 \gs {+}17) ( 
		112 \gl^2 {+} 4 \gs ( 5 \gs {+}38) {+} 
		8 \gl ( 12 \gs {+}43) {+}263)\ ,\\
		\mu_{17}&=36 ( 2 \gs {+}3) (4 \gl {+} 2 \gs {+}7 ) ( 12 \gl {+} 6 \gs {+}13) (  12 \gl {+} 
		6 \gs{+}17) ( 12 \gl {+} 6 \gs{+}19 )\ , \\
		\mu_{18}&=36 ( 4 \gl {+} 2 \gs{+}7) ( 12 \gl {+} 6 \gs {+}13) ( 12 \gl {+} 6 \gs {+}17) ( 
		12 \gl {+} 6 \gs {+}19) ( 12 \gl {+} 6 \gs {+}23)\ ,\\
		\mu_{19}&=-(73728 \gl^5{+} 288 \gs^5{+} 11952 \gs^4 {+} 127152 \gs^3{+} 552584 \gs^2{+}  1064386 \gs    {+} 18432 \gl^4 (7 \gs{+}32){+} 256 \gl^3 (342 \gs^2{+} 3555 \gs{+} 7417)\\
		&{+}16 \gl^2 (1800 \gs^3{+} 32508 \gs^2 {+} 146534 \gs {+}  190333 ) {+}	16 \gl (288 \gs^4 {+} 8136 \gs^3 {+}59596 \gs^2{+} 163350 \gs {+}151015){+}754803)\ ,\\
\end{aligned}
\end{equation*}
\end{multicols}
\begin{equation}
\begin{aligned}		
		\mu_{20}&=3 (12 \gl {+} 6 \gs{+}13) (9216 \gl^4 {+}144 \gs^4 {+}4032 \gs^3 {+}30072 \gs^2 {+} 85520 \gs {+} 1152 \gl^3 (12 \gs{+}55) {+} 
		16 \gl^2 (468 \gs^2 {+} 4824 \gs{+} 10307)\\
		& {+} 8 \gl (216 \gs^3{+} 3852 \gs^2{+} 17722 \gs {+} 23917){+}83097)\ ,\\
		\mu_{21}&=-3 (12 \gl {+} 6 \gs{+}13) (12 \gl {+} 6 \gs {+}17) (12 \gl {+} 6 \gs{+}19) (96 \gl^2{+} 12 \gs^2 {+} 160 \gs  {+} 4 \gl (18 \gs{+}91 ){+}357)\ ,\\
		\mu_{22}&=48 (12 \gl {+} 6 \gs{+} 13) (12 \gl {+} 6 \gs{+}17) (288 \gl^2{+}36 \gs^2 {+} 328 \gs  {+} 4 \gl (217 {+} 54 \gs){+}651)\ ,\\
		\mu_{23}&=-432 (4 \gl {+} 2 \gs{+}7) (12 \gl {+} 6 \gs{+}13) ( 12 \gl {+} 6 \gs {+}17) ( 12 \gl {+} 6 \gs {+}19)\ ,\\
		\mu_{24}&=3 ( 4 \gl {+} 2 \gs{+}7) ( 12 \gl {+} 6 \gs{+}13) (12 \gl {+} 6 \gs{+}17) (12 \gl {+} 6 \gs{+}19) (12 \gl {+} 6 \gs{+}23)\ .\\
		\end{aligned}
		\end{equation}

\section{Formulae for $\bf{BC_3}$}
\noindent
Explicitly, the integrals of motion read
\begin{equation}
\begin{aligned}
{I_4}&{\= p_1^4+\gs(\gs{-}1)\{p_1^2,\sfrac{1}{(x^1)^2}\}+\gs^2(\gs{-}1)^2\sfrac{1}{(x^1)^4}+2\gl(\gl{-}1)\textstyle \sum_{\ell\neq 1}^3 \bigl\{ p_1^2,  \sfrac{1}{(x^1-x^\ell)^2}{+}\sfrac{1}{(x^1+x^\ell)^2}\bigr\}}\\
&{-2 \im \gl(\gl{-}1)\textstyle \sum_{\ell\neq 1}^3 \{ p_1,  \sfrac{1}{(x^1-x^\ell)^3}{+}\sfrac{1}{(x^1+x^\ell)^3}\}+\gl^2(\gl{-}1)^2\textstyle \sum_{\ell\neq 1}^3\left( \sfrac{1}{(x^1-x^j)^4}{+}\sfrac{1}{(x^1+x^j)^4} \right)}\\
&{+4\gl(\gl{-}1)\left(\sfrac{1}{(x^1-x^2)^2}{-}\sfrac{1}{(x^1+x^2)^2}\right)p_1p_2+4\gl^2(\gl{-}1)^2\left(\sfrac{3}{((x^1)^2-(x^2)^2)^2}{+}4\sfrac{(x^1)^2+(x^2)^2}{((x^1)^2-(x^2)^2)^2}\sfrac{(x^1)^2+(x^3)^2}{((x^1)^2-(x^3)^2)^2}\right)}\\
&{+8 \gl(\gl{-}1)\gs(\gs{-}1) \left( \sfrac{1}{(x^1 x^2)^2}{+}\sfrac{2}{((x^1)^2-(x^2)^2)^2}\right)+\text{cyclic}\ .}
\end{aligned}
\end{equation}
\begin{equation}
\begin{aligned}
{I_6 }&=  \tfrac{1}{3}p_1^2 p_2^2 p_3^2+ \gs(\gs{-}1)\tfrac{1}{
	(x^1)^2 }p_2^2p_3^2-\gl(\gl{-}1)\bigl\{p_1p_2p_3^2,\sfrac{1}{(x^1-x^2)^2}{-}\sfrac{1}{(x^1+x^2)^2}\bigr\}+ \gs^2(\gs{-}1)^2\tfrac{1}{
	(x^1)^2(x^2)^2 }p_3^2  \\
	&+16\gl^2(\gl{-}1)^2\sfrac{(x^1x^2)^2}{((x^1)^2-(x^2)^2)^4}p_3^2
	+8\gs(\gs{-}1)\gl(\gl{-}1)\sfrac{1}{((x^1)^2-(x^2)^2)^2}p_3^2+16\gl^2(\gl{-}1)^2 \bigl\{p_1p_3, \sfrac{x^1x^3(x^2)^2}{((x^1)^2-(x^2)^2)^2((x^2)^2-(x^3)^2)^2} \bigr\} \\
&-4\gs(\gs{-}1)\gl(\gl{-}1) \bigl\{p_1p_3, \sfrac{x^1x^3}{(x^2)^2((x^3)^2-(x^1)^2)^2} \bigr\} -\gl^2(\gl{-}1)^2\sfrac{48 (x^1x^2)^2((x^1)^2+(x^2)^2)+160(x^1x^2x^3)^2}{((x^1)^2-(x^2)^2)^2((x^2)^2-(x^3)^2)^2((x^3)^2-(x^1)^2)^2}\\
&+\gs(\gs{-}1)\gl^2(\gl{-}1)^2\left( \sfrac{1}{(x^1)^2(x^2-x^3)^4}{+}\sfrac{1}{(x^1)^2(x^2+x^3)^4}{+} \sfrac{32(x^1)^2}{((x^1)^2-(x^2)^2)^2((x^3)^2-(x^1)^2)^2}\right) \\
&-2\gs(\gs{-}1)\gl(\gl{-}1)\bigl(\gl(\gl{-}1){-}4\gs(\gs{-}1)\bigr)\sfrac{1}{((x^1)^2((x^2)^2-(x^3)^2)^2}+\tfrac{1}{3}\tfrac{\gs^3(\gs{-}1)^3 }{
	(x^1)^2 (x^2)^2 (x^3)^2}+\text{cyclic}\ .
\end{aligned}
\end{equation}
\begin{equation}
\begin{aligned}
2{J_4}&=\{p_1^3,x^1\}+ \gs(\gs{-}1)\{p_1,\sfrac{1}{x^1}\}+\gl(\gl{-}1)\Bigl(6\textstyle\sum_{\ell\neq 1}\bigl(\sfrac{1}{(x^1-x^\ell)^2}+\sfrac{1}{(x^1+x^\ell)^2}\bigr)x^1p_1-\textstyle\sum_{\ell\neq 1}\bigl\{p_1,\sfrac{1}{x^1-x^\ell}+\sfrac{1}{x^1+x^\ell}\bigr\}\Bigr)\\
&+\text{cyclic}\ ,
\\
6{J_6}&=p_2^2p_3^2\{p_1,x^1\}{+}2\mi\gl(\gl{-}1)\Bigl(\sfrac{(p_1-p_2)}{(x^1-x^2)^3}+\sfrac{(p_1+p_2)}{(x^1+x^2)^3}\Bigr)\{p_3,x^3\}{+}\gs(\gs{-}1)p_3^2\Bigl(\{p_1,\sfrac{x^1}{(x^2)^2}\}+\{p_2,\sfrac{x^2}{(x^1)^2}\}\Bigr)\\
&-8\gl(\gl{-}1)\sfrac{(x^1x^2)^2}{((x^1)^2-(x^2)^2)^2}\bigr(\sfrac{p_1}{x^1}+\sfrac{p_2}{x^2}\bigr)p_3^2{-}16\gl(\gl{-}1)\tfrac{x^1 x^2 x^3}{((x^1)^2-(x^2)^2)^2}p_1 p_2 p_3{+}4\mi\gl(\gl{-}1)\sfrac{(x^1)^2+(x^2)^2}{((x^1)^2-(x^2)^2)^2}p_3^2 \\
&+8\mi\gl(\gl{-}1)\sfrac{x^1x^2}{((x^1)^2-(x^2)^2)^2}p_1p_2{+}\gs^2(\gs{-}1)^2\sfrac{1}{(x^1x^2)^2}\{p_3,x^3\}+8\gl(\gl{-}1)(2\gl^2{-}2\gl{-}9)\sfrac{(x^1x^2)^2}{((x^1)^2-(x^2)^2)^4}\{p_3,x^3\}\\
&-4\gl(\gl{-}1)\gs(\gs{-}1)\Bigl\{p_3,\bigr(\sfrac{(x^1)^2}{(x^2)^2((x^3)^2-(x^1)^2)^2}+\sfrac{(x^2)^2}{(x^1)^2((x^3)^2-(x^2)^2)^2}\bigr)x^3\Bigr\}+8\gl(\gl{-}1)\gs(\gs{-}1)\sfrac{1}{((x^1)^2-(x^2)^2)^2}\{p_3,x^3\}\\
&-12\gl(\gl{-}1)\sfrac{(x^1)^4+(x^2)^4}{((x^1)^2-(x^2)^2)^4}\{p_3,x^3\}+16\gl^2(\gl{-}1)^2\sfrac{(x^1x^2)^2}{((x^1)^2-(x^2)^2)^2}\Bigl\{p_3,\bigr(\sfrac{1}{((x^3)^2-(x^1)^2)^2}+\sfrac{1}{((x^2)^2-(x^3)^2)^2}\bigr)x^3\Bigr\} +\text{cyclic}\ .
\end{aligned}
\end{equation}

\newpage
\section*{}

\end{document}